# Theory of Programs

## Bertrand Meyer

While the theoretical study of programs fills volumes, a handful of concepts from elementary set theory suffice to establish a clear and practical basis.

Among the results:

- To describe a specification or a program, it suffices to define one relation and one set.
- To describe the concepts of programming, concurrent as well as sequential, three elementary operations on sets and relations suffice: union, composition and restriction.
- These techniques suffice to derive the axioms of classic papers on the "laws of programming" as straightforward consequences.
- To define both program correctness and refinement, the ordinary subset operator "⊆" suffices.

Paragraphs labeled "*Intuition*" relate the concepts to the experience of readers having done some programming. Readers familiar with other views of theoretical informatics will find comparisons in "*Comment*" paragraphs. Section 5 provides more discussion.

## 1 Programs

A program is a mathematical object: a constrained relation over a set of states.

*Definition*: program, specification, precondition, postcondition.

> A **program**, also known as a **specification**, over a state set $S$, consists of:
> - A relation $post: S \leftrightarrow S$, the program's **postcondition**.
> - A set $Pre \subseteq S$, the program's **precondition**.

*Notation*: $A \leftrightarrow B$ is the set of binary relations between $A$ and $B$, that is, $\mathbf{P}(A \times B)$. The domain of a relation $r$ is written $\underline{r}$ and its range $\overline{r}$.

*Intuition*: a program starts from a certain state and produces one of a set of possible states satisfying properties represented by *post*. *Pre* tells us which states are acceptable as initial states.

In the general case, more than one resulting state can meet the expectation expressed by *post*. Correspondingly, *post* is a relation rather than just a function.

The definition covers continuously running programs, such as those embedded in devices, since they are just repetitions of individual state transformations.

*Comment*: the usual view treats "program" and "specification" as distinct concepts, but all definitions of the purported difference are vague: for example, that a specification describes the "what" and a program the "how". The reason for the vagueness is that no *absolute* difference exists. An assignment is implementation to the application programmer and specification to the compiler writer. **Result**$^2 \cong$ *input* may look like a specification; but some "programming" languages accept it, letting the compiler derive a square-root algorithm. Any useful distinction must be *relative*: a program/specification *"specifies"* another. Section 3 will take advantage of this observation to introduce the notion of "contracted program", a *pair* of programs/specifications, one of which specifies the other. Until then, the two words are synonymous.

Particular choices for $S$ and for acceptable *post* and *Pre* determine particular styles of programming, such as the following.

*Definition*: deterministic, functional, imperative, object-oriented, object, procedural

---

A program $p$ is:

- **Deterministic** if $post_p$ is a function, and **non-deterministic** otherwise.
- **Functional** if every subset $C$ of $S$ is disjoint from $post_p(C)$, and **imperative** otherwise.
- **Object-oriented** if $S$ is of the form $1..n \to O$ for an integer $n$ and a set $O$ of "**objects**", and **procedural** otherwise.

---

*Notation*: For a relation $r$ in $A \leftrightarrow B$ and subsets $X$ of $A$ and $B$ respectively, $r(X)$ denotes the image of $X$, and $r^{-1}(Y)$ the inverse image of $Y$, by $r$. The relation is a "function" (short for "possibly partial function") if $r(\{x\})$, for any element $x$ of $A$, has at most one element. If it always has one, $r$ is "total". $A \to B$ is the subset of $A \leftrightarrow B$ containing total functions only. An integer interval is written $m..n$. Section 4.2 will present a more elaborate structure for $S$ in which the above characterizations apply to the "store" part.

$S_p$, $post_p$ and $Pre_p$ are the state set, postcondition and precondition of a program $p$. For the $i$-th program $p_i$ in a set of indexed programs we may use $S_i$, $post_i$ and $Pre_i$.

The principal concepts of programming, studied in the rest of this presentation, are independent of such choices of style and of the properties of $S$.

*Definition*: feasibility

---

A program $p$ is **feasible** if $Pre_p \subseteq \underline{post_p}$.

---

*Intuition*: $Pre_p$ tells us when we *may* apply the program, and $post_p$ what kind of result it *must* then give us. A program/specification is safe for us to use if it meets its obligation whenever we meet ours. Feasibility expresses this property: for any input state satisfying $Pre_p$, at least one output state satisfies $post_p$.

*Comment*: it is possible to avoid introducing feasibility as a separate condition: define the concept of program by *post* only, and just *define Pre* as *post*. Then every program is feasible. This model, however, does not fit the practice of programming. Often we are given a general relation (such as **Result**$^2 \cong$ *input*) that is not satisfiable for every possible input state; we must find an input domain (such as *input* $\geq 0$) on which it is. Hence the definition of "program" as the general concept and "feasible program" as a desirable special case.

*Definition*: program equivalence

> Two programs are **equivalent** if they have the same *Pre* and the same *post / Pre*.

> *Notation*: For a relation $r$ and subsets $X$ and $Y$ of its source and target sets, $r/X$ and $r \setminus Y$ are $r$ restricted to the domain $X$ (meaning $r \cap (X \times S)$) and corestricted to the codomain $Y$ (meaning $r \cap (S \times Y)$). Two straightforward properties (restriction and corestriction theorems) are that $r/X \subseteq X$ and $r \setminus Y \subseteq Y$.

*Intuition*: the results of a program only matter for input states satisfying the precondition.

*Definition*: refines, specifies, abstracts

> A program/specification $p_2$ **refines** another, $p_1$, and $p_1$ **specifies** (or **abstracts**) $p_2$, if:
> P1   $S_2 \supseteq S_1$         -- Extension
> P2   $Pre_2 \supseteq Pre_1$     -- Weakening
> P3   $post_2 \subseteq_{Pre_1} post_1$    -- Strengthening

> *Notation*: $r \subseteq_X r'$ means $(r/X) \subseteq r'$; in other words, whenever $r$ maps an element of $X$ to a result, $r'$ maps it to the same result. The same conventions applies to other operators on relations, as in $r =_X r'$. Note the names (extension, weakening, strengthening) associated with the three conditions of the definition.

*Intuition*: a refinement of $p$ gives more detail than $p$, but still satisfies all properties of $p$ relevant to users of $p$. So it must cover all of $p$'s states, accept all the input states $p$ accepts and, for these states, only yield results that $p$ could also yield. It may have more states, a more tolerant precondition, and yield only some of the results that $p$ could yield (in particular, reduce non-determinism).

*Comment*: in practice, we might want a refined program to work on a different set of states. Then $S_1$ would *map* to a subset of $S_2$, rather than *being* that subset (P1). It is possible to generalize the notion of refinement in this spirit.

*Theorem*: Refinement Theorem

> P4   Refinement is a preorder.

*Proof*: Since $\supseteq$ is an order relation, reflexivity, antisymmetry and transitivity hold for the program's state set and precondition parts. As to the postcondition part, reflexivity is trivial; for transitivity, if $Y \supseteq X$ then $r_3 \subseteq_Y r_2 \subseteq_X r_1$ implies $r_3 \subseteq_X r_1$. Refinement is not antisymmetric, but if $p_1$ refines $p_2$ and $p_2$ refines $p_1$ it follows from P2 that $Pre_1$ is identical to $Pre_2$ and then from P3 that $post_1$ and $post_2$ coincide on $Pre_1$, so the programs are equivalent according to the definition above. In other words, refinement is an order relation if we consider equivalent programs as equal.

*Notation*: P3 and the refinement theorem justify writing "$p_2$ refines $p_1$" as $p_2 \subseteq p_1$. This will be a general convention: given some operator § on relations, we extend it to programs so that $p_2 \S p_1$ means $post_2 \S post_1$, with a suitable condition on preconditions. More examples appear below.

*Definition*: implementation

> An **implementation** of $p$ is a feasible refinement of $p$.

*Intuition*: not every refinement of a specification is feasible. For example the infeasible program <∅, S> refines every specification over *S*. Hence the importance of finding *feasible* refinements, also known as implementations. This concept still does not provide a distinction between programs and specifications.

*Notation*: For a known set of states *S*, <*post, Pre*> is the program of postcondition *post* and precondition *Pre*.

*Theorem*: Implementation Theorem

| P5 | A specification/program having an implementation is feasible. |

*Intuition*: the statement — if a specification has a feasible refinement, it is itself feasible — seems obvious in light of the words it uses, but in fact requires a proof.

*Proof*: Let *p* be the specification and *i* the implementation; we must prove that $Pre_p \subseteq \underline{post_p}$. Weakening (P2) tells us that $Pre_p \subseteq Pre_i$, and feasibility of *i* that $Pre_i \subseteq \underline{post_i}$. Hence property A: $Pre_p \subseteq \underline{post_i}$. Strengthening tells us that $post_i \underset{Pre_p}{\subseteq} post_p$, hence property B: $\underline{post_i} \cap Pre_p \subseteq \underline{post_p}$. From A and B we deduce that $Pre_p \subseteq \underline{post_p}$.

# 2 Operations on specifications and programs

The fundamental operations of elementary set theory yield fundamental operations on specifications and programs:

- Union gives choice (intersection, for its part, does not have a directly useful application).
- Restriction gives conditionals.
- Composition of relations gives sequence ("compound" or "block" in programming languages).
- Composition combined with union for symmetry gives concurrency (parallelism).
- Composition of a relation with itself a variable number of times (power) gives loops.

The following definitions cover all these programming constructs and some others. Only the first three (those of 2.1) refer directly to the basic concepts defined so far; all the rest follow as combinations of those three.

## 2.1 Basic constructs

*Definition*: choice, composition, restriction

| Name | Notation | Mathematical definition | | Programming intuition |
|---|---|---|---|---|
| | | **Postcondition** | **Precondition** | |
| Choice (or: union) | $p_1 \cup p_2$ (Dijkstra: $p_1 \;[]\; p_2$) | $post_1 \cup post_2$ | $Pre_1 \cup Pre_2$ | Performs like $p_1$ or like $p_2$ |
| Composition (or: sequence, compound, block) | $p_1 \,;\, p_2$ | $(post_1 \setminus Pre_2)\,;\, post_2$ | $Pre_1 \cap post_1^{-1}(Pre_2)$ | Performs first like $p_1$ then like $p_2$ |
| Restriction (guarded command) | $C\colon p$ (Dijkstra: $C \to p$) | $post_p / C$ | $Pre_p$ | Performs like $p$ on $C$ |

*Notation*: In the "postcondition" column, the semicolon ";" denotes composition of functions or relations, in the order of application, so that $(r \,;\, s)\,(X)$ is $s\,(r\,(X))$. (Mathematical texts often use $s \circ r$ for $r \,;\, s$.) "Dijkstra" means the notation of [3].

*Comment*: the first two operators transpose well-known mathematical operations, union in the first case and composition in the second, to programs. They consequently retain their symbols, "$\cup$" and ";". No confusion results since it is always clear whether the operands are sets (including relations) or programs.

*Comment*: in the definition of program composition, it might seem sufficient to use $post_1 \,;\, post_2$ for the postcondition (rather than $(post_1 \setminus Pre_2) \,;\, post_2$); but that approach is incorrect because $post_1$ could pass on to $post_2$ some elements that do not satisfy $Pre_2$. An example (with $S$ a set of integers) is $p_1 = <\{[1, 1], [1, 2]\}, \{1\}>$ and $p_2 = <\{[1, 1], [2, 2]\}, \{1\}>$; here $post_1 \,;\, post_2$ is $\{[1, 1], [1, 2]\}$, but results from applying $post_2$ to 2, not part of its precondition. At first sight the precondition $Pre_1 \cap post_1^{-1}\,(Pre_2)$ appears to guard against this risk, but it does not: this precondition guarantees that $p_1$ yields *at least one element* satisfying $Pre_2$, but does not stop $p_1$ from also yielding *other* results that do not satisfy $Pre_2$. (Underlying this discussion is a mathematical property of the image operator: $r\,(r^{-1}\,(C)) \supseteq C$, a superset property only, not an equality.) We will see that invariant preservation (2.7) also requires the corestriction to $Pre_2$. (Instead of corestriction we may use restriction: $(post_1 \setminus Pre_2) \,;\, post_2$ is the same as $post_1 \,;\, (post_2 \,/\, Pre_2)$.)

*Theorem*

| | |
|---|---|
| P6 | For feasible operands and arbitrary conditions, the above operators yield feasible programs. |

*Proof*: the definition of feasibility is $Pre_p \subseteq \underline{post_p}$. For choice, we note that for relations $r_1$ and $r_2$ $\underline{r_1 \cup r_2} = \underline{r_1} \cup \underline{r_2}$; for composition, that $\underline{r_1 \,;\, r_2} = \underline{r_1} \cap (r_1^{-1}\,(\underline{r_2}))$ (for $r_1 \,;\, r_2$ to be applicable to an element $x$, $r_1$ must be applicable to $x$ and yield from $x$ at least one element to which $r_2$ is applicable); for restriction, that $\underline{r \,/\, C} = \underline{r} \cap C$.

*Theorems*

Properties of the basic operators directly reflect those of their mathematical counterparts. Choice, like union of sets, is commutative; composition of programs, like composition of relations, is not. Choice and composition are associative, so we may apply them without parentheses to any number of operands, as in $p_1 \,;\, p_2 \,;\, \ldots \,;\, p_n$. In addition:

| | | | | |
|---|---|---|---|---|
| P7  | $C_1\colon (C_2\colon p)$ | $=$ | $C_2\colon (C_1\colon p)$ | -- Restriction is commutative. In fact: |
| P8  | $C_1\colon (C_2\colon p)$ | $=$ | $(C_1 \cap C_2)\colon p$ | |
| P9  | $C\colon (p_1 \cup p_2)$ | $=$ | $(C\colon p_1) \cup (C\colon p_2)$ | -- Restriction distributes over choice. |
| P10 | $C\colon (p_1 \,;\, p_2)$ | $=$ | $(C\colon p_1) \,;\, p_2$ | -- Composition absorbs restriction. |
| P11 | $q \,;\, (p_1 \cup p_2)$ | $=$ | $(q \,;\, p_1) \cup (q \,;\, p_2)$ | -- Composition distributes left… |
| P12 | $(p_1 \cup p_2) \,;\, q$ | $=$ | $(p_1 \,;\, q) \cup (p_2 \,;\, q)$ | -- … and right over choice. |

(Choice, however, does not distribute over composition.) The proofs are straightforward but must cover both postcondition and precondition.

The following special programs are of interest:

- $\langle \emptyset, \emptyset \rangle$, called *Fail*: nowhere applicable. (Sometimes known as "abort" or "halt".)
- $\langle S \times S, S \rangle$, called *Havoc*: always applicable, but we may assume nothing about the result.
- $\langle (\lambda x: S \mid \{x\}), S \rangle$, called *Skip*: always applicable, leaves the state exactly as it was.

   *Notation*: generalized lambda notation serves to define relations in $A \leftrightarrow B$, using either $\lambda x: A \mid Y$ where $Y$ is a subset of $B$ (as here for *Skip*), or $\lambda x_1: A; x_2: B \mid p(x_1, x_2)$ where $p$ is a two-variable predicate. A program/specification is **total** if its precondition is $S$.

*Comment*: all three are feasible; in the case of *Fail* note the difference with the infeasible example $\langle \emptyset, S \rangle$ used after the definition of "implementation". *Skip* and *Havoc* are total.

*Theorems*

| | | | | |
|---|---|---|---|---|
| P13 | $(p\ ;\ Skip)$ | $= (Skip\ ;\ p)$ | $= p$ | |
| P14 | $(p \cup Fail)$ | $= (Fail \cup p)$ | $= p$ | -- Does not hold in the demonic theory. |
| P15 | $(Fail\ ;\ p)$ | $= (p\ ;\ Fail)$ | $= Fail$ | |
| P16 | $(p \cup Havoc)$ | $= (Havoc \cup p)$ | $= Havoc$ | |
| P17 | $(p\ ;\ Havoc)$ | $= (Pre_p: Havoc)$ | | |
| P18 | $p$ | $\subseteq (C: p)$ | | -- Reminder: "$\subseteq$" on programs is refinement. |
| P19 | If $D \subseteq C$, then $(C: p) \subseteq (D: p)$. | | | -- Order reversal (precondition weakening). |
| P20 | If $q \subseteq p$, then $(C: q) \subseteq (C: p)$. | | | -- Refinement safety, see below. |
| P21 | If $q_1 \subseteq p_1$ and $q_2 \subseteq p_2$, then $(q_1 \cup q_2) \subseteq (p_1 \cup p2)$ and $(q_1\ ;\ q_2) \subseteq (p_1\ ;\ p_2)$. | | | |
| P22 | $p \subseteq (Pre_p: Havoc)$ for any $p$. | | | |
| P23 | $p \subseteq Havoc$ for any total $p$. | | | |
| P24 | $p \subseteq Fail$ if and only if $p = Fail$ | | | -- *Fail* is refined only by itself… |
| P25 | $Fail \subseteq p$ if and only if $p = Fail$ | | | -- … and refines only itself. |

*Comment* (varieties of non-determinism): $p_1$ does not generally refine $p_1 \cup p_2$ because of the precondition $Pre_1 \cup Pre_2$. "Internal choice", which has the same postcondition as choice but the precondition $Pre_1 \cap Pre_2$, satisfies refinement but not P11, distributivity over composition. (Consider $q = \langle \{[0, 1], [0, 2]\}, \{0\} \rangle$, $p_1 = \langle \{[1, 0]\}, \{1\} \rangle$, $p_2 = \langle \{[2, 0]\}, \{2\} \rangle$: under internal choice the precondition is empty for the left side of P11 and $\{1\}$ for the right side.).

Another terminology is that choice is "angelic" and internal choice can be "demonic". The theory of programs has a demonic sister, obtained by choosing internal choice for all the operator definitions that rely on choice. The discussion will point out places where the difference matters.

*Notation*: "$\cup$" for choice is a new example (after "$\subseteq$" for refinement and ";" for composition) of extending set operators to programs. The following application of this idea is also useful:

| Name | Notation | Postcondition | Precondition | Programming intuition |
|---|---|---|---|---|
| Corestriction | $p \setminus C$ | $post_p \setminus C$ | $Pre_p \cap post_p^{-1}(C)$ | $p$, applied only when results satisfy $C$ |

(On the other hand we do not need a restriction notation $p \,/\, C$ since we already have $C: p$.)

The first of the following properties shows that corestriction can be defined from restriction and composition.

*Theorems*

| | | | |
|---|---|---|---|
| P26 | $(p \setminus C)$ | $= (p \,;\, (C: Skip))$ | |
| P27 | $(p_1 \cup p_2) \setminus C$ | $= (p_1 \setminus C) \cup (p_2 \setminus C)$ | -- Compare with P9. |
| P28 | $(p_1 \,;\, p_2) \setminus C$ | $= p_1 \,;\, (p_2 \setminus C)$ | -- Compare with P10. |
| P29 | $(p \setminus C)$ | $\subseteq C$ | -- Refinement. Compare with P18. |
| P30 | If $D \subseteq C$, then $(p \setminus D)$ | $\subseteq (p \setminus C)$ | -- Compare with P19. |

The restriction and corestriction theorems apply to programs: $\underline{C: p} \subseteq C$ and $\overline{p \setminus C} \subseteq C$.

*Notation*: in the same spirit, the range and domain notations apply to programs: $\underline{p}$ is a synonym for $Pre_p$; and (more importantly) $\overline{p}$ is a synonym for $post_p \,(\underline{p})$, the set of values that $p$ can actually yield.

Properties P20 and P21 extend to all well-behaved operators in the following sense.

*Definition*: refinement safety

| An operator § on programs is **refinement-safe** if $q_1 \subseteq p_1$ and $q_2 \subseteq p_2$ implies $(q_1 \,§\, q_2) \subseteq (p_1 \,§\, p_2)$. |
|---|

*Counter-examples*: intersection of programs, defined as intersecting both postconditions and preconditions, is not refinement-safe: with a set of integers for $S$, $\{0\}$ for all preconditions, and postconditions $\{[0, 0], [0, 1]\}$ for $p_1$ and $p_2$, $\{[0,0]\}$ for $q_1$ and $\{[0,1]\}$ for $q_2$, the conditions of the definition are met, but $q_1 \cap q_2$ has an empty postcondition and hence does not refine $p_1 \cap p_2$, which is just $p_1$. Another counter-example is program difference (set difference of postconditions, intersection of preconditions). The theory of programs, however, eschews such operators:

*Theorem*: refinement safety

| P31 All the operators on programs introduced in this article are refinement-safe. |
|---|

In a corresponding sense, the program properties "functional" and "object-oriented" are refinement-safe (but not their opposites, "imperative" and "procedural").

## 2.2 Atomic concurrency

Composition, while associative, is not commutative: when we combine existing programs or specifications, it forces us to decide in which order we want them to perform. If you find this obligation irksome, you need concurrency. Concurrent combination (in its "atomic" form) is sequential composition made symmetric through association with its commutative colleague, choice.

*Definition*: concurrency

| Name | Notation | Definition | Programming intuition |
|---|---|---|---|
| Atomic concurrency | $p_1 \,\|\, p_2$ | $(p_1 \,;\, p_2) \cup (p_2 \,;\, p_1)$ | Performs once like each of $p_1$ and $p_2$ |

*Theorems*: properties of atomic concurrency

| P32 | Atomic concurrency "$\|$" is commutative, associative and refinement-safe. | | |
|---|---|---|---|
| P33 | $p_1 \| (p_2 \cup p_3)$ | $= (p_1 \| p_2) \cup (p_1 \| p_3)$ | -- Concurrency distributes over choice, left… |
| P34 | $(p_1 \cup p_2) \| p_3$ | $= (p_1 \| p_3) \cup (p_2 \| p_3)$ | -- … and right. |
| P35 | $C: (p_1 \| p_2)$ | $= (C: p_1) \| (C: p_2)$ | -- Restriction distributes over concurrency… |
| P36 | $(p_1 \| p_2) \setminus C$ | $= (p_1 \setminus C) \| (p_2 \setminus C)$ | -- … and so does corestriction. |
| P37 | $(p_1 \,;\, p_2)$ | $\subseteq (p_1 \| p_2)$ | -- Sequential composition refines concurrency… |
| P38 | $(p_2 \,;\, p_1)$ | $\subseteq (p_1 \| p_2)$ | -- … in any order. |

Concurrency generally does not refine composition, but in one particular case it does.

*Definition*: commuting programs

Two specifications/programs commute if $(p_1 \,;\, p_2) = (p_2 \,;\, p_1)$.

*Example and counter-example*: if $S$ is the set of functions $PERSON \to \mathbb{Z}$, recording people's bank account balances, consider a set of programs indexed by persons $p$ and integers $n$: the postconditions of $deposit_{p,n}$ and $withdraw_{p,n}$ express that the output differs from the input only by having the balance of $p$ respectively increased and decreased by $n$. All these programs commute with each other. They do not commute, however, with the program $reset_p$ setting $p$'s balance to zero.

*Theorem*

| P39 | If $p_1$ and $p_2$ commute, then $(p_1 \| p_2) = (p_1 \,;\, p_2)$. |
|---|---|

(Not just refinement, but equality. Immediate generalization to more than two programs.)

*Intuition*: commuting programs are a boon for concurrent computation, since they open up many possible realizations for "computing" program results (finding values satisfying $post_p$) on actual "computers" (the physical devices that ensure postconditions). Assume for example a large number of *deposit* and *withdraw* operations with various clients and amounts. If the specification is that at the end of the trading day the balance of each should be correct (initial, plus accumulated deposits, minus accumulated withdrawals), any distribution of the operations among any number of computers in any order is suitable. In such cases concurrency is an optimization mechanism.

*Comment*: Commuting is not refinement-safe: with $S = \{0, 1\}$, the total programs of postconditions $\lambda x \mid \{0\}$ and $\lambda x \mid \{1\}$ both refine *Havoc*, which commutes with itself, but do not commute with each other since composing them in both orders yields for $0$ the values $0$ and $1$ respectively. Abstraction (the inverse of refinement) also does not preserve commuting: *Skip* and *Havoc* do not commute even though *Skip* commutes with itself and refines both.

Refinement and abstraction do preserve commuting for deterministic programs with identical preconditions, but are also of limited interest in this case.

## 2.3 Non-atomic concurrency

The atomic concurrency operator has a fixed level of granularity, defined by its operands: if they are themselves complex programs built out of simpler components, it will not interleave these components. For example let *on* be "switch on the light", *off* "switch it off" and $p$ "say whether the light is on". Assuming that in the initial state the light is on, $(on \,;\, off) \parallel p$ will always say no, regardless of which of the operands of the "$\parallel$" goes first, since $(on \,;\, off)$ is equivalent to *Skip*.

The practice of concurrency often calls for finer-grain control on concurrency. Here you might want $p$ to execute at the beginning, in the middle (between *on* and *off*), or at the end. Such flexibility causes much of the difficulty of concurrent programming, since it opens up the possibility of "data races" (inconsistent orderings of operations, in some executions only); but a general theory of programming must provide a model for it, given here by a ternary operator.

| Name | Notation | Definition | Programming intuition |
|---|---|---|---|
| Non-atomic concurrency | $(p_1, p_2) \parallel q$ | $((p_1 \parallel q) \,;\, p_2) \cup$ $(p_1 \,;\, (p_2 \parallel q))$ | Performs once like each operand, with $p_1$ before $p_2$ |

*Notation*: the only new symbol is the comma, used at a place where the semicolon of composition could also appear. The reuse of "$\parallel$" is only for convenience: the above "Notation" entry describes a new three-operand operator. Its "Definition" entry relies on the previously defined atomic concurrency operator "$\parallel$". No confusion arises since the non-atomic operator only occurs in conjunction with the comma.

*Comment*: we do need a specific operator, because proposing a distributive-style law involving standard composition ";" would raise inconsistencies. For example, $(on \,;\, off) \parallel p$ cannot give any other result than *Skip* $\parallel p$; if you want to allow interleaving, you should specify a finer level of granularity, as in $(on \,,\, off) \parallel p$. In the first case the atomic unit of concurrency on the left side is $(on \,;\, off)$; in the second case there are two atomic units, *on* and *off*.

Non-atomic concurrency is associative on its first two operands $p_1$ and $p_2$, so you may use commas to separate any number of program operands of non-atomic concurrency. (Reduced to one operand, as in $(p_1) \parallel q$, atomic and non-atomic "$\parallel$" coincide, as they should for consistency.) One may also put $q$ first, writing $q \parallel (p_1, p_2)$. In other words, the notation lets us use a comma, to specify a finer granularity of interleaving, where we might otherwise use a semicolon.

*Theorems*

| | | | |
|---|---|---|---|
| P40 | $(p_1, p_2) \parallel q$ | $= (q \,;\, p_1 \,;\, p_2) \cup (p_1 \,;\, q \,;\, p_2) \cup (p_1 \,;\, p_2 \,;\, q)$ | |
| P41 | $(p_1 \,;\, p_2) \parallel q$ | $\subseteq (p_1, p_2) \parallel q$ | -- Coarser-grained refines finer-grained. |
| P42 | $p_1 \,;\, (p_2 \parallel q)$ | $\subseteq (p_1, p_2) \parallel q$ | -- First "law of exchange" of [8]. |
| P43 | $(p \parallel q_1) \,;\, q_2$ | $\subseteq p \parallel (q_1, q_2)$ | -- Second "law of exchange" of [8]. |

*Proof* of P42: the left side is $(p_1 \,;\, p_2 \,;\, q) \cup (p_1 \,;\, q \,;\, p_2)$, which from P40 (itself a direct consequence of the definition) is a subset of the right side; similarly for P43. Both of these properties appear in [8] as fundamental axioms of concurrency, but here they are simple theorems.

It is straightforward to symmetrize the non-atomic concurrency notation to $(p_1, p_2) \parallel (q_1, q_2)$, yielding the generalized law of exchange from [8]: $(p_1 \parallel q_1) \,;\, (p_2 \parallel q_2) \subseteq (p_1, p_2) \parallel (q_1, q_2)$.

## 2.4 Conditionals

*Definition*: conditionals

| Name | Notation | Definition | Programming intuition |
|---|---|---|---|
| Guarded conditional (Dijkstra) | **if** $C_1$: $p_1$ **[]** $C_2$: $p_2$ **end** | $(C_1: p_1) \cup (C_2: p_2)$ | Performs like $p_1$ on $C_1$ or like $p_2$ on $C_2$ |
| If-then-else | **if** $C$ **then** $p_1$ **else** $p_2$ **end** | $(C: p_1) \cup (C': p_2)$ | Performs like $p_1$ on $C$ and like $p_2$ elsewhere |

*Notation*: $C'$, for a subset $C$ of $S$, is its complement: $S - C$. The usual programming notation is "**not** $C$" (see 2.5 below). The guarded conditional is in fact not new since **if** $q_1$ **[]** $q_2$ **end** was introduced in 2.1 as a synonym for $q_1 \cup q_2$, but it highlights the important case of $q_1$ and $q_2$ being restrictions.

*Theorems*: conditional instruction properties

| | |
|---|---|
| P44 | The guarded conditional is commutative. |
| P45 | Both forms of conditional are associative. |
| P46 | Both forms of conditional distribute over choice and concurrency. |
| P47 | If $D_1 \subseteq C_1$ and $D_2 \subseteq C_2$, then (**if** $D_1$: $p$ **[]** $D_2$: $q$ **end**) $\subseteq$ (**if** $C_1$: $p$ **[]** $C_2$: $q$ **end**). |
| P48 | If $q_1 \subseteq p_1$ and $q_2 \subseteq p_2$, then (**if** $C$: $q_1$ **[]** $C$: $q_2$ **end**) $\subseteq$ (**if** $C$: $p_1$ **[]** $C$: $p_2$ **end**). |
| P49 | If $q_1 \subseteq p_1$ and $q_2 \subseteq p_2$, then (**if** $C$ **then** $q_1$ **else** $q_2$ **end**) $\subseteq$ (**if** $C$ **then** $p_1$ **else** $p_2$ **end**). |
| P50 | (**if** $C$ **then** $p_1$ **else** $p_2$ **end**) $=$ (**if** $C'$ **then** $p_2$ **else** $p_1$ **end**) |
| P51 | $(C: p)$ $=$ (**if** $C$: $p$ **end**) |
| P52 | (**if** $C_1$: $p_1$ **[]** $C_2$: $p_2$ **end**) $\subseteq$ $C_1$: $p_1$   -- A conditional refines any of its branches |
| P53 | $(D:$ (**if** $C_1$: $p$ **[]** $C_2$: $q$ **end**)) $=$ (**if** $(D \cap C_1)$: $p$ **[]** $(D \cap C_2)$: $q$ **end**) -- Distributivity. |
| P54 | (**if** $C$ **then** $p_1$ **else** $p_2$ **end**) $=$ (**if** $C$: $p_1$ **[]** $C'$: $p_2$ **end**) |
| P55 | (**if** $C$ **then** $p_1$ **else** $p_2$ **end**) $=$ (**if** $C'$ **then** $p_2$ **else** $p_1$ **end**) |

*Proof*: Refinement properties follow from P19, P21 and other earlier theorems; for P53, see P8 and P9.

*Comment*: As seen next, "$\cap$" in these rules can also be written "**and**". On P44, note that if-then-else is not commutative (but see P50), and on P46 that conditionals do not distribute over composition.

*Notation*: As a result of associativity (P45), a conditional of either kind can be applied to more than two operands. If-then-else uses **elseif** for the second to next-to-last branches, as in **if** $C_1$ **then** $p_1$ **elseif** $C_2$ **then** $p_2$ **else** $p_3$ **end**. A conditional can also take just one operand: for the guarded conditional, **if** $C$: $p$ **end** is the same as $C$: $p$; for if-then-else, by convention, **if** $C$ **then** $p$ **end** is an abbreviation for **if** $C$ **then** $p$ **else** *Skip* **end**.

## 2.5 Conditions

Two special conditions are useful for building programs. *True* is another name for $S$, and *False* another name for the empty set. They should not be confused with the similarly named constants of propositional calculus: *True* and *False* are, like all conditions, sets (subsets of $S$). In fact the theory of programs relies on elementary set theory rather than directly on logic.

It is easy, however, to define boolean-like operators on conditions: **and** and **or** as other names for "∩" and "∪", **not** as another name for complement (in P55 we may write *C'* as **not** *C*), **implies** or "⇒" as other names for "⊆", and so on. Here, in addition to P19, are some properties involving operations on conditionals.

*Theorems*

| | | | |
|---|---|---|---|
| P56 | (*True*: *p*) | = | *p* |
| P57 | (*False*: *p*) | = | *Fail* |
| P58 | *p* \ *True* | = | *p*      -- Here "\" is corestriction on programs. |
| P59 | *p* \ *False* | = | *Fail* |
| P60 | (**if** *True* **then** $p_1$ **else** $p_2$ **end**) | = | $p_1$ |
| P61 | (**if** *False* **then** $p_1$ **else** $p_2$ **end**) | = | $p_2$      -- And similarly for guarded conditionals. |
| P62 | **and**, **or**, **not**, **implies** distribute over choice, restriction and conditionals. | | |

*Proof*: for P59, note that the postcondition of *p* \ *False* is $post_p \cap (S \times False)$, that is, an empty relation (since *False* is the empty set).

## 2.6 Loop

*Definition*: repetition constructs

| Name | Notation | Definition | Programming intuition |
|---|---|---|---|
| Fixed repetition | $p^i$ for any natural integer *i* | $p^0 = \underline{p}$: *Skip* $p^{i+1} = (p \,;\, p^i)$ | *p* repeated *i* times |
| Arbitrary repetition | **loop** *p* **end** (or $p^*$) | $\bigcup_{i \geq 0} p^i$ | *p* repeated any number of times |
| "While loop" | **from** *a* **until** *C* **loop** *b* **end** (or *a*; **while not** *C* **loop** *b* **end**) | *a* ; (**loop** *C'*: *b* **end**) \ *C* | *a*, then *p* repeated until *C* holds |

*Notation*: in the definition of the while loop, it does not matter how we parenthesize the "\"; see P28. Since composition is associative, the inductive expression for fixed repetition can also be written ($p^i; p$).

*Intuition*: **loop** *p* **end** is the program that performs like *p* repeated some finite (but unknown) number of times. Cyclic programs, such as those embedded in devices, follow this pattern. The rest of the present discussion concentrates on the **from** *a* **until** *C* **loop** *b* **end** loop, which starts like *a* then performs like *b*, the loop's "body", as many times as needed (possibly zero) until reaching a state satisfying *C*. In slightly different terms: for the loop to yield a result from a given input state *x*, that result must be one of the first elements of *C* reached by successive executions of *b* after *a*. All the previous states are not in *C*, so they are in *C'*, meaning that what we are iterating is not the whole *b* but just *C'*: *b*.

From distributivity follows another expression of the loop:

*Theorem*: Loop Lemma

> P63  The loop $l =$ **from** $a$ **until** $C$ **loop** $b$ **end** can be written $\underset{i \geq 0}{\cup}\, q_i$, where $q_i$ is $a\,;\, (C'\!:\!b)^i \setminus C$.
> As a consequence, $\overline{l} = \cup\, \overline{q_i}$.

   *Notation reminder*: $\overline{p}$, a subset of $\overline{post_p}$, is the set of values that $p$ can produce.

*Intuition*: $q_i$ represents a restricted version of the loop, which yields a result (satisfying $C$) after exactly $i$ iterations. The loop is the union of all its partial versions.

*Comment*: unlike previous constructs, the loop does not automatically get feasibility from the feasibility of its operands: it is possible for $a$ and $b$ to be feasible while $l$ is not. (A trivial example is **from** *Skip* **until** *False* **loop** *Skip* **end**, for which every $q_i$ is *Fail*.) A loop is feasible if and only if for every suitable state $s$ there exists an integer $i$ such that $(a\,;\, (C'\!:\!b)^i)\,(\{s\})$ contains an element in $C$; in other words, that $q_i(\{s\})$ is not empty.

  The feasibility condition for loops relies on the notion of invariant.

## 2.7 Invariants

*Definition*: invariant

> A condition $I$ is an **invariant** of a program/specification $p$ if $post_p(I \cap \underline{p}) \subseteq I$.

*Intuition*: an invariant is called that way because if it holds before application of $p$ it will hold afterwards. More precisely, for the initial condition we need not the whole of $I$ but just $I \cap \underline{p}$, since results of $p$ only matter when $p$ starts in its precondition. The following two theorems result directly from the definition.

*Theorems*

> P64  Any $I$ disjoint from $\underline{p}$ is an invariant of $p$.
> P65  If $I$ and $J$ are invariants of $p$, so are $I \cup J$ and $I \cap J$.

*Comment*: image properties involving intersection are usually not as strong as those involving union, because $r(I \cap J)$ is only a subset of $r(I) \cap r(J)$, rather than equal to it as with "$\cup$"; but P65 has both operators on an equal footing.

*Theorem*: Invariant Refinement Theorem

> P66  If $I$ is an invariant of $p_1$ and $p_2 \sqsubseteq p_1$, then $I$ is an invariant of $p_2\,/\,Pre_1$.

*Comment*: in practice, the precondition often stays the same under refinement, but in the general case $p_2$ might have a broader precondition; there is no guarantee that the original invariant will hold for the new states, hence the restriction to $Pre_1$.

*Definition*: invariant-preserving operator

> An operator on programs is **invariant-preserving** if any invariant of all its program operands is also an invariant of the operator's result.

*Example*: program composition is invariant-preserving.

*Proof*: assume $I$ is an invariant of both $p_1$ and $p_2$. The definition of program composition (2.1) gives $(post_1 \setminus Pre_2)\ ;\ post_2$ as the postcondition of $q = (p_1\ ;\ p_2)$. From P26 and properties of image $((r_1\ ;\ r_2)\ (A) = r_2\ (r_1\ (A)))$ and restriction $((C\colon r)\ (A) = r\ (C \cap A))$, it follows that $post_q\ (q \cap I) = post_2\ (Pre_2 \cap Res_1)$ where $Res_1 = post_1\ (q \cap I)$. Since $I$ is an invariant of $p_1$, $Res_1 \subseteq I$; since it is also an invariant of $p_2$, then, $post_2\ (Pre_2 \cap Res_1) \subseteq I$.

*Comment*: the discussion after the definition of program composition in 2.1 noted that taking just $post_1\ ;\ post_2$ as postcondition for $p_1\ ;\ p_2$ would not yield a feasible result: we need the corestriction to $Pre_2$. This property is also essential for invariant preservation: without it we would be applying $post_2$ not to $Pre_2 \cap Res_1$ but just to $Res_1$, on which $post_2$ does not preserve the invariant.

This result about composition is only a particular case of the following general property.

*Theorem*: General Invariant Theorem

> P67   All the program operators defined so far are invariant-preserving.

*Proof*: the result for all the basic operators (choice, sequence, restriction) follows from the set-theoretical properties of relational image, including the following in addition to those used in the preceding proof: $r\ (C \cup D) = r\ (C) \cup r\ (D)$; $r\ (C \cap D) \subseteq r\ (C) \cap r\ (D)$; $(r \setminus D)\ (C) \subseteq r\ (C)$. The subsequent operators (concurrency, conditional) are defined from the basic ones and retain their invariant preservation.

Every element of the infinite unions that define loops is made out of basic operators and, by induction, is invariant-preserving. Since union maintains this property, the loops themselves possess it. They benefit, however, from a more specific form of the notion of invariant.

*Definition*: loop invariant

> A **loop invariant** of **from** $a$ **until** $C$ **loop** $b$ **end** is a subset of $\overline{a}$ that is an invariant of $C'\colon b$.

The Invariant Refinement Theorem, P66, implies that a "loop invariant" is an "invariant", in the general sense, of the part of the loop that comes after initialization ($a$). The following theorem yields a stronger form of the relationship between the two concepts.

*Theorem*: Loop Correctness Theorem

> P68   If $I$ is a loop invariant of the loop $l = ($**from** $a$ **until** $C$ **loop** $b$ **end**$)$, then $\overline{l} \subseteq C \cap I$

*Intuition*: The theorem characterizes the fundamental property of loops [11] [5]: the goal of a loop is to obtain on exit ($\overline{l}$) a combination of the exit condition ($C$) and a judiciously chosen invariant ($I$, a weakening of the desired result).

*Proof*: since $I$ is an invariant of $C'\colon b$, it is an invariant of $(C'\colon b)^i$ for any integer $i$; since $I$ is also a subset of $\overline{a}$, it follows that $\overline{q_i} \subseteq I$ for every $i$, with $q_i$ as defined in the Loop Lemma, P63. Then, from the second part of the Loop Lemma, $\overline{l} \subseteq I$. In addition, the corestriction theorem tells us that $\overline{q_i} \subseteq C$ as well, again for every $i$; this property extends to $\overline{l}$.

*Comment*: despite its fundamental role, the Loop Correctness Theorem does not fully cover the theory of loops because it says nothing about feasibility. It states that loop results — elements of $\overline{l}$ — possess interesting properties, but not that such elements exist for every legal input state. In fact, a loop yielding no results at all (an empty $\overline{l}$) would satisfy the theorem. In the traditional terminology of theoretical informatics, the theorem is a "*partial correctness*" result, useful only if we can also guarantee "*termination*". The complementary theorem follows.

*Theorem*: Loop Feasibility Theorem

> P69   For feasible $a$ and $b$, the loop **from** $a$ **until** $C$ **loop** $b$ **end** is feasible if both:
> - $\underline{b} \cup C$ is a loop invariant.
> - $C'\colon post_b$ is well-founded.

*Notation*: a "well-founded" (or "Noetherian") relation is one that admits no infinite chain.

*Proof*: For an arbitrary element $s$ of $\underline{a}$, define $S_0$ as $a(\{s\})$ and $S_{i+1}$ as $(C'\colon post_b)(S_i)$. In other words, $S_i$ is $(a\,;\,(C'\colon b)^i)(\{s\})$, the result of iterating the loop $i$ times on $s$. Then $q_i(\{s\}) = S_i \cap C$. We will prove a non-disjointness property: $S_i$ and $C$ cannot be disjoint for all $i$. Then there exists an $i$ such that $q_i(\{s\})$ contains at least one element, which is in $\overline{q_i}$ and hence in $\overline{l}$, showing that the loop is feasible.

   The proof of non-disjointness is by contradiction. Assume that $S_i$ and $C$ are disjoint for all $i$. By induction, $S_i$ is not empty: since $a$ is feasible, $S_0$ is not empty; and if $S_i$ is not empty, the invariant property tells us that $S_i \subseteq \underline{b} \cup C$; with $S_i$ disjoint from $C$ this means $S_i \subseteq (\underline{b} \cap C')$ which implies, $b$ being feasible, that $S_{i+1}$, the image of $S_i$ by $C'\colon post_b$, is not empty. But then elements of successive non-empty sets in the infinite sequence $S_i$ are related by $C'\colon post_b$, an impossibility since the relation is well-founded.

*Comment*: while the theorem gives a general condition for loop feasibility, it is often not practical to check directly that $C'\colon post_b$, the loop body, is well-founded. A standard technique is to map states to a simpler domain on which it is easier to check that the counterpart of $post_b$ is well-founded, according to the following definition.

*Definition*: loop variant

> A **loop variant** of **from** $a$ **until** $C$ **loop** $b$ **end** is a total function $v$ from $S$ to a set $V$ equipped with a well-founded relation "$<$", such that $v(s') < v(s)$ for any $s$ in $C'$ and $s'$ in $post_b(s)$.

The existence of a variant shows that $post_b$ itself is well-founded, fulfilling the second condition of the Loop Feasibility Theorem. The most frequent choice for $V$ is the set of natural integers and for "$<$" the usual order relation on integers.

# 3  Contracted programs

There is, as noted, no difference of principle between specifications and programs. Since this is not the conventional view, let us see if we can find a place for a meaningful distinction.

We already saw that the first attempt, stating that specifications are abstract and programs concrete, does not make the cut, since "level of abstraction" is a relative notion (the example was an assignment instruction, abstract for some and concrete for others). A seemingly more promising intuition is that programs are *executable* while specifications are descriptive; but it does not work either, since executability is also a relative notion, which has evolved through the history of computing (the example here was **Result**$^2$ = *input*, which can be executable in a high-level language).

The relevant criterion is correctness. As captured by the notion of feasibility, a specification can be inconsistent (if it tells you that the result must be zero and also that it can be one) or consistent; but it makes no sense to ask whether it is correct. Correct with respect to what? Probably with respect to the customers' desires, or to their actual needs, but these would have to be written down as another, higher-level specification, only pushing the problem further. We do know, however, what it means for a program to be correct: it performs according to a stated specification. Correctness is a relative notion.

Indeed what truly distinguishes a program from a specification, in the common usage of these terms, is neither the level of abstraction nor the possibility of execution, but the existence of *two* programs/specifications in the sense of the present theory, such that one of them is a refinement of the other. The following notation reflects this analysis.

*Definition*: contracted program, specification part, contract, implementation part, correctness

> The notation **require** *Pre* **do** *b* **ensure** *post* **end**, a **contracted program**, asserts that *b* is an implementation of the specification/program *p* = <*post*, *Pre*>.
>
> Then *p* is the **specification part**, or **contract**, and *b* the **implementation part**. The contracted program is also said to be a **correct program**.

*Reminder*: an implementation of *p* is a feasible refinement of *p*. The refinement theorem, P5, indicates that *p* is feasible as well. The definition of refinement indicates that the precondition of *b* is a superset of *Pre* and its postcondition a subset of *post*. (The name *b* stands for "body".)

*Intuition and comment*: the notion of contracted program simply introduces a programming notation for the concept of refinement. Since a program is useless without a precise understanding of what it is supposed to do, program authors should only produce contracted programs. Regrettably, this practice is not yet universal.

The above definition provides a final clarification of what programs in the usual sense of the term (*contracted programs* in the present theory) really are: **a program is a proof obligation**. Writing **require** *Pre* **do** *b* **ensure** *post* **end** is a way to state that *b* must refine *p*, and requires the author, before clicking "Run", to click "Verify". Section 5.2 expands on this definition.

*Theorem*

> P70 If *post* ⊆ *post'*, *Pre'* ⊆ *Pre*, and **require** *Pre* **do** *b* **ensure** *post* **end** is a contracted program, so is **require** *Pre'* **do** *b* **ensure** *post'* **end**.

*Comment*: in this case, since we keep the implementation and go to a new specification, we can only strengthen the precondition and weaken the postcondition.

The following concepts are defined for given *Pre*, *post* and *b*.

*Definitions*: weakest precondition, strongest postcondition

> $post_b$ */ Pre*, also written *b* **sp** *Pre*, is the **strongest postcondition** of *b* for *Pre*.
> $b - post_b - post$, also written *b* **wp** *post*, is the **weakest precondition** of *b* for *post*.

*Intuition*: $post_b - post$ is the set difference of two relations, giving us the set of pairs that belong to the first but not to the second. Its domain, $post_b - post$, is the set of states for which *b* produces at least one result that *post* could never produce. Subtracting this domain from *b*, the domain of *b*, gives us the set of states on which *b* is guaranteed to agree with *post*.

The following property justifies the terms "strongest" and "weakest".

*Theorem*

> P71 If **require** *Pre* **do** *b* **ensure** *post* **end** is a correct program, then (*b* **sp** *Pre*) ⊆ *post* and *Pre* ⊆ (*b* **wp** *post*).

*Proof*: Let *p* be <*post, Pre*>. Since *b* is a refinement of *p*, $post_b \subseteq_{Pre} post$ by the definition of refinement, yielding the first property of the theorem. By refinement, *Pre* ⊆ *b*; the just mentioned property $post_b \subseteq_{Pre} post$ implies that $post_b - post$ is disjoint from *Pre*, so $Pre \subseteq b - post_b - post$, giving us the second property.

As a corollary, we get a compact definition of program correctness.

*Theorem*

> P72 **require** *Pre* **do** *b* **ensure** *post* **end** is correct if and only if $Pre \subseteq b - post_b - post$.

*Theorems*

| | | | |
|---|---|---|---|
| P73 | *b* **sp** *False* | = | *Fail* |
| P74 | *b* **wp** *Fail* | = | *False* |
| P75 | *Fail* **sp** *C* | = | *Fail* |
| P76 | *Fail* **wp** *p* | = | *False* |
| P77 | *b* **sp** (*p* ∪ *q*) | = | (*b* **sp** *p*) ∪ (*b* **sp** *q*) |
| P78 | *b* **wp** (*p* ∪ *q*) | ⊇ | (*b* **wp** *p*) ∪ (*b* **wp** *q*) |

(and so on). As an example of why P78 is not an equality, consider postconditions $\{[0, 1], [0, 2]\}$ for $b$, $\{[0, 1]\}$ for $p$ and $\{[0, 2]\}$ for $q$, all with precondition $\{0\}$. Then both $b$ **wp** $p$ and $b$ **wp** $q$ are empty (since $b \text{---} p$ has postcondition $\{[0, 2]\}$ and $b \text{---} q$ has $\{[0, 1]\}$), but $b$ **wp** $(p \cup q)$ is $\{0\}$. This property is related to the comment (after P25) that in the angelic theory $p_1$ does not generally refine $p_1 \cup p_2$.

*Definition*: generalizing refinement to contracted programs

> If $q \sqsubseteq p$ ($q$ refines $p$), **require** *Pre* **do** $q$ **ensure** *post* **end** refines **require** *Pre* **do** $p$ **ensure** *post* **end**.

*Comment*: it is possible to generalize the definition further by having different *specification* parts.

*Definition and theorem*: Most Abstract Implementation

> P79   For feasible $p$, **require** $p$ **do** $p$ **ensure** $post_p$ **end**, the **most abstract implementation** of $p$, is a correct program, which every implementation of $p$ refines.

*Intuition*: The most abstract implementation is the specification used as its own implementation.

# 4 States and environments

The exact nature of $S$, the state set, varies considerably between application domains and the formalisms supporting programming (*programming languages* as defined next in section 5). Some properties, however, are common to most variants.

## 4.1 Mappings

The state tracks the evolution, during the computation, of certain elements of information relevant to the results. As a consequence, a state almost always includes (as its essential components) one or more mappings between these elements and their current values. "Mapping" is a general term roughly equivalent to "function"; in programming, since the memories of both humans and computers are finite, these functions will also be finite. $S$, then, includes components of the form *Name* $\twoheadrightarrow$ *Value* for appropriate sets of names and values.

> *Notation*: $A \rightarrowtail B$ is the set of possibly partial functions, and $A \twoheadrightarrow B$ the set of finite functions, from $A$ to $B$. Inclusions are: $(A \twoheadrightarrow B) \subseteq (A \rightarrowtail B) \subseteq (A \leftrightarrow B)$ and $(A \rightarrow B) \subseteq (A \rightarrowtail B)$.

## 4.2 Environment and store

It is common for the state to have two clearly identified components: the environment and the store, also known as the static and dynamic parts. In a simple case, with a set *Var* (for "variables") of names and a set *Type* representing the types of possible values, the environment is of the form *Var* $\twoheadrightarrow$ *Type* and the store of the form *Var* $\twoheadrightarrow$ *Value*. This division reflects the typical process of executing programs on a computer:

- A first step known as **compilation** creates the environment.
- The actual computation, known as **execution**, takes place in the second step, which builds and transforms the store, constrained by environment built in the first step.

One of the advantages of this approach is that it requires programmers to define types for every variable, making it possible to detect mistakes (such as applying a boolean operation to integer variables) in the first step; in that case the second step does not take place until the programmer has corrected the mistake. Such a process limits the risk of erroneous computation. Another advantage is that it is not necessary to repeat the first step once it has succeeded: subsequent executions of the same program, applied to different input states, can use the result of the compilation.

*Definitions*: declaration, instruction

> A function in $S \nrightarrow S$ is a **declaration** if it leaves the store part unchanged, and an **instruction** if it leaves the environment part unchanged.

*Intuition*: it is good practice to separate the two kinds of operation; declarations set up the environment; instructions, working in a defined environment, change only the store.

*Comment*: the characterization of programming styles (functional, object-oriented) in section 1 properly applies to the store component of the state. So do the definitions of *Skip* and *Fail* (2.1) if we wish to treat these operations as instructions.

## 4.3 Notational principles: cartesian product considered harmful

The preceding discussion has stopped short of specifying $S$ as the cartesian product $E \times M$ where $E$ is the environment and $M$ (for "memory") the store. It does not even use the common programming-like "record" notation (*environment*: $E$; *store*: $M$) (mathematically denoting a function in $Tag \nrightarrow U$, where $Tag$ is the set of names to the left of the colons and $U$ the union of the sets to their right, with the constraint that the function's values for the $i$-th tag are in the $i$-th set). The two models are isomorphic and either one would be suitable for a purely mathematical discussion, but for modeling software concepts they are too constraining.

The reason is that the theory of programs, like the development of programs, calls for more incremental notations, allowing us to extend and adapt existing models. Both cartesian product and the record notation are closed: if you have defined a concept such as "state" through a particular set of components, such as the environment and the store, and later want to add a component, you must rework all previously defined operations (functions or relations) on states. An example of such an operation is a declaration, defined as $\lambda\, e, m \mid [d\,(e), m]$ where $d$ is an function on the environment (for example, if $e$ is or includes a mapping in $Var \nrightarrow Type$, $d$ yields a new version of the mapping, extended with a new pair such as [$n$, *INTEGER*]). If you add a third component to the concept of state, this definition, which yields a pair rather than a triple, no longer makes sense.

Cartesian product is not the only culprit: definition by alternation is just as bad. It is common to use definitions of the form $L \triangleq J \mid K$, specifying that an element of $L$ is disjointly either an element of $J$ or an element of $K$. (Again there is a simple mathematical model, applicable even if $J$ and $K$ are not disjoint: the notation describes pairs in $\{1, 2\} \times (J \cup K)$ such that the second element is in $J$ for 1 and in $K$ for 2, with generalization to any number of alternatives.) This notation suffers from the same drawback: adding an alternate breaks all previous derivations.

In programming, the "object-oriented" method of programming, with its concept of "inheritance", is an effective remedy to these problems. The theoretical side requires such solutions too.

This article does not introduce the details of the appropriate notation but it is useful to see the principal convention, used as the replacement for cartesian product. When a set needs to be defined with a number of components, we give each a name, as in

>   **S component**
>       *environment*: *E*
>       *store*: *M*

This mathematical notation simply asserts the existence of two total functions, *environment* in $S \to E$ and *store* in $S \to M$. Projections are written (for a state *s*) *s*.*environment* and *s*.*store*. A function on composite objects defined from functions on their components is of the form

>   **on S update**
>       *environment'* = *d* (*environment*)
>   **end**

denoting a function in $S \to S$, with the rule that the function leaves unchanged any component not named, here *store*; the example denotes the function that for any state of components *e* and *m* yields the state of components *d* (*e*) and *m*. In the end, these notations are equivalent to the cartesian product and record forms, but there is a practical difference: you can include as many "**component**" and "**on**" definitions as you like, even for the same target set *S*, in an incremental fashion. In many cases, the existing specifications can remain as they were, thanks to the rule that not citing a component in an **on** ... **update** ... specification means leaving it unchanged.

Such definitions are cumulative: mathematically, the resulting specification is the cartesian product of all the **component** declarations. (This convention assumes that the graph of declarations involves no recursion; it can be extended through fixpoint techniques to support recursive definitions.) A similar convention applies to sets defined by alternation.

In both cases, a simple notation supports "lifting" an operation on a component into an operation on the whole. For example if *d* is an operation on the environment it is convenient to treat it also as an operation on the state, which as in the above **on** ... **update** ... example leaves all other components unchanged.

This article will not need further details of these techniques, but they are important for the practical development of specifications and programs.

### 4.4 Kinds of state

While the precondition is a set of states, the postcondition in the general case is a relation over two states, initial and final; a common term is "*two-state assertion*". For example, we may want to specify that the initial state contains a positive number (precondition, a set) and the final state its approximate square root (postcondition, a relation between input and output).

Some postconditions, for example "the output is positive", do not involve the initial state:

*Definition*: Markovian, one-state

> A postcondition *post* is **Markovian**, or **one-state**, if $\forall\ s, s_1, s_2 \mid (s_1\ \textbf{post}\ s) = (s_2\ \textbf{post}\ s)$

> Notation: $x\ \mathbf{r}\ y$, for a relation $r$, expresses that $[x, y] \in r$ (the relation connects the two elements). The equality in the definition is equivalence (equality between two boolean properties).

*Intuition*: a Markovian postcondition characterizes only the final state, regardless of the input.

*Comment*: a useful program produces different results for different inputs, and so is generally not Markovian if considered as a whole. But postconditions are often expressed as intersections (conjunctions) of properties, some of which can be Markovian; for example the result's square is close to the input *and* the result is non-negative. The Markovian property can also characterize intermediate steps in the program. This observation extends to the following state properties.

*Definition*: trivial, irrelevant, relevant

> For a postcondition *post*, a state *s* is:
> - **Trivial** if $\forall\ s_1 \mid s\ \textbf{post}\ s_1$.
> - **Irrelevant** if $\forall\ s_1, s_2 \mid (s\ \textbf{post}\ s_1) = (s\ \textbf{post}\ s_2)$.
> - **Relevant** if not irrelevant ($\exists\ s_1, s_2 \mid (s\ \textbf{post}\ s_1) \neq (s\ \textbf{post}\ s_2)$).

*Intuition*: if a state is trivial, a transition to *any* other state will fulfill the postcondition. If it is irrelevant, it plays no role in whether the next state satisfies the postcondition.

*Theorem*

> P80  A specification <*post*, *Pre*> is feasible if and only if every state in *Pre* is either trivial or relevant.

*Proof*: We may assume a non-empty *Pre*. ($\Rightarrow$) Assume the specification is feasible. If a state *s* in *Pre* is irrelevant and not trivial, $s\ \textbf{post}\ s_1$ holds for no $s_1$. Feasibility implies $Pre \subseteq \overline{post}$, meaning there is an $s_1$ such that $s\ \textbf{post}\ s_1$, yielding a contradiction. ($\Leftarrow$) Assume every state $s$ in $\overline{Pre}$ is either trivial or relevant. If *s* is trivial, it is in $\overline{post}$. If *s* is relevant, then there exist $s_1$ and $s_2$ so that either $s\ \textbf{post}\ s_1$ or $s\ \textbf{post}\ s_2$, so it is also in $\overline{post}$.

# 5  Languages and programming

"Programming" is the act of writing contracted programs according to the preceding definitions. Such a program has two parts: the contract represents the goal of the program, as advertised to its users; the implementation represents the operations that will run on the computer. The definition ensures that the implementation matches the contract.

## 5.1 Programming languages

If the contract is given, in the form of *Pre* and *post*, programming consists of solving **require** *Pre* **do** *b* **ensure** *post* **end**, viewed as an equation of which *b* is the unknown.

The Most Abstract Implementation, as defined above, yields a trivial solution, often non-deterministic, to the equation: $post_b = post\,/\,Pre$, $Pre_b = Pre$. The reason why that solution is generally of little use, and programming an interesting endeavor, is the *practical* difference between contract and implementation. For $b$ we seek a relation $post_b$ that a material computer can process (not necessarily directly, but through the services of tools such as "compilers"). For the specification, since the goal is to describe the problem, we can rely on a broader set of mathematical mechanisms.

In both cases we need a repertoire of mathematical tools to build programs and specifications.

*Definition*: programming language, specification language

> A **programming language** over a state set $S$, also known as a **specification language** over $S$, is a set of feasible programs over $S$. In practice it is given by:.
> - A finite set of base programs, obtained from a finite set of relations in $S \leftrightarrow S$ (serving as base postconditions) and a finite set of subsets of $S$ (serving as base preconditions).
> - A finite set of operators for deriving new correct programs from previously defined ones.

*Intuition*: a programming language is a set of possible programs. Any useful programming language is infinite, but it is derived from a few basic postconditions and preconditions, and a few operators to combine them. Many of these basic elements, introduced in the earlier sections of this presentation, can be used by programming languages regardless of the application domain:

- *Havoc*, *Skip* and *Fail* as base programs, plus *True* and *False* ($S$ and $\varnothing$) as base preconditions.
- The program construction operators of section 2, including the three basic ones (choice, composition and restriction) and those derived from them (concurrency, conditionals, loops).

Beyond these universal elements, a language will offer specific mechanisms for the intended application domain, beginning with a suitable set $S$ of states and a suitable set of operations over $S$.

Since specification and implementation are often considered separate activities, it is frequent to find separate specification and programming languages. Another approach is to use a single language; this approach is in fact required if we want to produce *correct* programs (contracted programs), which include both a contract and an implementation. (As noted after the definition of refinement, it is possible to define a variant of the theory in which the state set changes under refinement, but at the price of much added complexity.) Approaches to producing reliable software can be hampered by a failure to understand the fundamental unity of the programming process: in spite of the obvious differences in levels of abstraction, the problems and solutions, for which this presentation offers a mathematical framework, are the same. (Reference [13] discusses the *seamlessness* of the development process in a software-engineering rather than mathematical context, and [14] develops its application to software requirements.)

Absent such a single framework, not only is it hard to produce correct software; even *expressing* what it means for the program to be correct is a challenge, since the implementation and specification belong to different worlds (such as an ordinary programming language and some specification framework). One must define a mapping between these two worlds, potentially adding complexity and supplementary correctness issues.

With a single *S* and a single specification and programming language, the language description will identify, among the language's mechanism, the subset suitable for implementation. Then the requirement on program authors is simply to produce a final version **require** *Pre* **do** *b* **ensure** *post* **end** of the program in which the implementation part *b* only relies on that subset. Establishing correctness means establishing:

- Refinement: $b \subseteq \,$<*post, Pre*>.
- Feasibility: *Pre* $\subseteq$ *post* (or alternatively, thanks to the implementation theorem P5, $Pre_b \subseteq post_b$ if the preceding condition holds).

One can express these properties convincingly, and prove them, since all three components, *post*, *Pre* and *b*, are part of the same mathematical framework, even if the last one restricts itself to a subset of that framework's mechanisms.

## 5.2 Approaches to programming

The most common approach to programming today ignores the *Pre* and *post* elements of the definition, concentrating only on building implementations *b* from a programming language with the hope that in some informal sense they will match the corresponding user needs. We may call this the "hacking approach"; it has little to commend itself if correctness is part of the objectives.

At the other extreme, a "refinement approach" [17] [1] [15] has made its mark in informatics research and led to such development methods as B. If we set out to implement a given contract, the Most Abstract Implementation theorem P79 tells us that we may use the contract itself — specifically, <*post, Pre*> — as its own first implementation. Refinement as a software development method starts with this first version and repeatedly takes advantage of theorems to choose a "refinement" in the sense of the formal definition, P2 and P3, of the previous implementation until reaching an implementation that belongs to the implementation part of the language.

This approach is elegant but faces some obstacles:

- *Hindsight*: we seldom know the entire specification in advance. This uncertainty is not necessarily a mark of incompetent software engineering: the very process of implementation suggests new elements of specification — "esprit de l'escalier" as discussed in [13].
- *Extendibility*: even if the specification is initially clear, it usually changes as a project progresses and after initial deliveries. If a change affects a property that was used in an early step of the refinement process, it becomes necessary to redo much of the work. (*Invariants*, which play an important role in refinement methods, can help control this change process [2].)
- *Reusability*: A top-down refinement process does not easily take into account implementations previously produced for variants or subsets of the problem. It is desirable for a development process to accommodate a bottom-up component, supporting reuse.

The ideal process should combine the best elements of the "hacking" and "refinement" approaches, retaining the practicality of the first and the rigor of the second. It is not the goal of the present discussion to present such a process, but a general definition helps set the stage.

*Definition*: programming

> **Programming** is the process of devising interesting contract-implementation pairs and discharging the associated proof obligations.

The starting point for any step in the process may indifferently be:

- A contract element, for which we have to devise a satisfactory implementation (top-down).
- Existing implementation elements (bottom-up). Ideally these elements already have full contracts. In practice, they often have no contracts, or incomplete ones; part of the process then involves a form of reverse engineering: uncovering the precise intent of the components and writing the contracts.

This approach seems to yield the necessary flexibility while accommodating the need for rigor and proofs. It yields a useful view of programs.

*Slogan*: program

> **Program = Contract + Implementation + Proof obligation**

# 6 Discussion

This article applies to programming the standard method on which science and engineering rely to solve practical problems in any application domain:

- Develop a mathematical model resulting in equations (in the present case, the feasibility equation *Pre* $\subseteq$ *post* and the program equation **require** *Pre* **do** *b* **ensure** *post* **end**, where *b* is the unknown).
- Solve the equation.
- Build the solution in the application domain.

The main argument for the model developed in the preceding sections is the simplicity of its premises: the mathematical baggage is elementary set theory, learned in high school around the age of 15. The construction relies on just three mechanisms from that theory: union, composition and restriction. The approach seems to have the potential to cover all the relevant concepts of programming, although the present article takes only a first dig.

## 6.1 Axioms or theorems?

In theoretical informatics the habit has often been different: devising axiomatic theories. The most developed example is the admirable work of Hoare and colleagues [7][8]. A notable property of these efforts is that they postulate their laws; then "*of course, the mathematician should also design a model of the language, to check completeness and consistency of the laws, to provide a framework for the specifications of programs, and for proofs of correctness*" [7]. The justification for this method — postulate your ideal laws, the model will follow — is that it has, in Russell's words cited in [9], "*the advantages of theft over honest toil*".

However good the wisecrack, this is not how normal mathematics works. Unless your last name is Euclid or Peano, or your first name Alfred or Bertrand (and even in this last case, only if you inherited a peerage), few people will pay attention to axioms you assert on them as if walking down from Mount Sinai. Imagine a world where every mathematical concept were defined axiomatically; in trigonometry, sine and cosine would be postulated as functions satisfying certain properties — the sum of their squares is 1, the derivative of the former is the latter, and so on; and similarly for every important notion. People would quickly tire of having to make incessant leaps of faith.

We expect instead, when presented with new results, to see them *derived*, in the form of definitions and theorems, from what we already know. True, it is often a mark of elegance, for the presenter of a theory and of its laws, to prove that the theory is the simplest possible construction satisfying the laws; but it is a mark of politeness to perform this feat only as a bonus step, coming after an explanation relying only on material already familiar to the reader.

Stretching Russell's aphorism, we may note that even if Balzac's observation ("*The secret of great fortunes without apparent cause is a forgotten crime*") may explain the *origin* of some hereditary peerages, just as axioms explain the foundations of mathematics, in practice most hereditary peers find it less bothersome to obtain the objects of their daily desires through "honest toil", or at least honest means, than by stealing.

These observations do not rule out occasional reliance on the axiomatic method in the introduction of theories. Aphorisms aside, however, it is hard to justify asserting properties as postulates when they can be proved as theorems. When a manageable mathematical derivation from known concepts exists, it should be the first choice.

As the presentation of the theory of programs has attempted to show, such exactly is the situation with programming. Programs are just relations over sets. An informal and non-exhaustive review of the axioms of classic articles such as [7] and its extension to concurrency [8] (not considering properties specific to individual calculi), as well as [6] and [10], suggests that the properties they introduce can be derived, often straightforwardly, from the framework of this article; many indeed appear above as theorems.

Many authors seem to have a suspicion, conscious or not, of the set-theoretical basis of programming; but most — an important exception is Hehner with his "predicative programming" [6] — resist the obvious solution of explicitly building the theory on that basis. They prefer to posit axioms, even if these axioms mimic the elementary properties of set operators. An example is the seminal "Laws of Programming" article [7] (with its recent extension to concurrency [8]), whose authors axiomatically introduce operators with names such as "$\cup$" for non-deterministic choice and "$\subseteq$" for refinement. They never suggest that these could actually *be* the standard mathematical operators bearing the same names; but they cover several pages of *Communications of the ACM* with such fascinating "axioms" as $P \cup (Q \cup R) = (P \cup Q) \cup R$. One wonders whether the thought ever arose that if it associates like union, commutes like union, distributes like union, and typographically uses the exact symbol of union, perhaps it is union.

## 6.2 Keeping simple things simple

Because informatics already struggles to describe inherently complex phenomena, we should not introduce complexity of our own making. Programming theory does not always keep the complexity of the descriptions commensurate with the complexity of the described. Another seminal paper of great elegance [10] introduces the "natural semantics" of the if-then-else conditional thus:

$$\frac{\rho \models (E_2 \Rightarrow \alpha)}{\rho \models (\textbf{if } \textit{True} \textbf{ then } E_2 \textbf{ else } E_3 \textbf{ end}) \Rightarrow \alpha}$$

with a similar rule for the *False* case. In words: if in the environment $\rho$ the expression $E_2$ evaluates to $\alpha$, then in $\rho$ the expression **if** *True* **then** $E_2$ **else** $E_3$ **end** also does. The companion rule tells us that if $E_3$ evaluates to $\beta$ the expression with *False* instead of *True* evaluates to $\beta$.

In reality, if-then-else is a very simple concept. It expresses that one may solve a problem by partitioning the domain into two parts and using a different solution in each. Euler would undoubtedly have explained it to his 15-year-old princess pupil [4] by a little illustration:

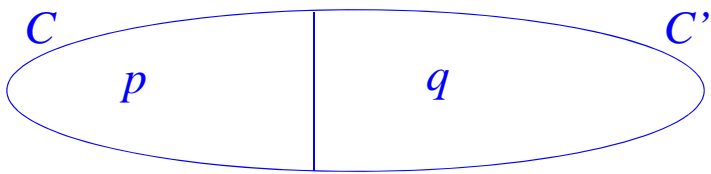

and she would have understood on the spot. (A pedagogical presentation of the theory of programs' concepts should indeed use Euler-Venn diagrams throughout, although this article has shunned them under the presumption that its putative audience does not need pictures.)

Instead, the above "natural" semantics refers to advanced concepts of mathematical logic and notions such as the "environment" ($\rho$), which are a distraction from the idea of a conditional instruction. These observations do not put into question the value of [10] and other classic semantic articles, which were conceived as research advances, not tutorials. But they highlight the benefit, as a domain gets understood better, of seeking simplicity and trimming down the set of prerequisite concepts to the indispensable.

## 6.3 De-emphasizing the program text

One source of complication in theories of programming is reverence for the program text.

Almost every discussion of programming — saying "almost" just to be on the safe side — starts by defining a programming language. (Denotational or operational semantics often starts with *two* languages, one to express programs and the other to express their meanings.)

This attitude seems to be a leftover from the early days when parsing was the difficult problem. Programmers and theorists were awe-struck when Backus, Bauer, Hopper and others showed that instead of coding with zeros and ones it was possible to use a human-readable notation and have it translated automatically. The program text became the alpha and omega of programming. But it is only an artifact. A computer is a mathematical machine for computing pairs in relations. All the rest is decoration.

Programming is no more about programs than electricity is about plugs.

Parsing is the original informatics problem and even though it has long lost its theoretical difficulty it remains our unconscious template for all others. Semantic specification, for example, often looks like a smarter kind of parsing, also starting from program texts and deriving its properties — just more interesting properties. Denotational semantics defines "meaning functions" operating on program texts. Electrical engineers, if they worked that way, would start from plugs, dutifully noting how different Swiss, French and Italian plugs are from each other. In reality, of course, what counts is the electrical current — the same in all three countries, with their interconnected networks — and specifically the relevant equations.

In programming too a more productive approach — the application to semantics of the idea of *unparsing*, the reverse of parsing — is to start from an analysis of what we need mathematically: what kinds of postconditions and preconditions give rise to useful specifications and realistic implementations. From this analysis we construct programming notations, not the other way around. For example we do not start from if-then-else as a given construct of interest, but identify the union of two relations as a relevant concept. We consequently derive suitable notations to express it, each adapted to different mathematical situations: if the relations' domains are provably disjoint, **if** $C$ **then** $p$ **else** $q$ **end**; otherwise, the guarded conditional **if** $C$: $p$ [] $D$: $q$ **end**. The notation for contracted programs, **require** *Pre* **do** $b$ **ensure** *post* **end**, continues in this spirit: it is simply a way to express the mathematical property that $b$ implements *<post, Pre>*.

Far from lessening the value of the traditional objects of interest in informatics, such as programs and programming languages, this reversal of perspective makes them even more interesting, turning them from arbitrary products of taste and circumstance into rationally justified modes of expression for useful mathematical concepts.

## 6.4 The basic duality

The presentation of the theory has highlighted a characteristic property of programming: the natural need for two distinct methods to assess what a program can do and whether it will actually get to do it. This separation is hardly a revelation: in theoretical discussions of programming it recurs under many guises, such as partial correctness versus termination, safety versus liveness, loop invariants versus loop variants. The present discussion provides more evidence of its inevitability. Note the two loop theorems (Loop Correctness, P68, and Loop Feasibility, P69) and the separate definitions of "program" and "feasible program". Even the attempt to define "correct program" in a single formula, P72, requires two operands reflecting the two sides of the question.

In [3] Dijkstra also attempted to cover loops through a single rule, but in practice one must still separately use an invariant and a variant. Partly blessing, partly curse, the duality seems to be an inescapable part of informatics, reflecting built-in limits of human reason.

## 7 Perspectives

The thesis of this article is that it is possible to found all of programming on a small set of concepts from elementary set theory. The discussion has shown the basic applications, but is only a start. (Also note that the theorems have not been mechanically checked.) Future tasks include:

- Reconstructing entire programming languages on that basis.
- Using the theory to build a "Formal Language Innovation Platform" (FLIP) for experimenting with programming language mechanisms.
- Developing it towards specific approaches to programming, particularly object-oriented.
- Assessing whether the approach can produce effective program verification tools.
- Assessing whether it can help teach programming, including at the elementary level.

## 8 Acknowledgments

The authors invoked explicitly or not in section 6 (Hoare and coauthors, Kahn, Dijkstra, Scott/Strachey/Plotkin and other pioneers of denotational semantics), complemented by Abrial for his work on Z and B and by Mills and Gries, deserve deep acknowledgments for pioneering the formal approach to programs and programming. Back's and Morgan's seminal work on refinement (following Wirth's) is another fundamental inspiration. Hehner's work on Predicative Programming is a comprehensive theory of programming based on binary relations, corresponding to the postconditions of the present work. (I am also indebted to him for a particularly careful reading of the first draft.) Also influential have been informal comments by David Parnas on the merits of different assertion styles. A note by Shaoying Liu [16], criticizing a purported deficiency in classical refinement approaches (the risk of refining into an unfeasible program), suggested the need for a proper notion of feasibility.

I am grateful to Daniel de Carvalho and Colin Adams for corrections on the first draft.

## 9 References

[1] Back, refinement papers.

[2] Michael Butler, personal communication.

[3] Dijkstra, *A Discipline of Programming*.

[4] Euler, *Lettres à une Princesse d'Allemagne sur divers Sujets de Physique et de Philosophie*, 1760-1762.

[5] Furia, Meyer, Velder, *Computing Surveys* invariant article.

[6] Hehner, *Predicative Programming*

[7] Hoare, original paper on Laws of Programming.

[8] Hoare and van Staden, newer article.

[9] Hoare and van Staden, slides accompanying [8].

[10] Kahn, *Natural Semantics*.

[11] Meyer, IFIP 1980 paper.

[12] Meyer, ETL.

[13] Meyer, OOSC.

[14] Meyer, *Multirequirements*.

[15] Morgan, *Programming from Specifications*.

[16] Shaoying Liu, paper and slides from the 2014 Futatsugi Festschrift.

[17] Wirth, stepwise refinement.